\documentclass{article}
\usepackage{amsmath}
\usepackage{graphicx}
\usepackage{color}
\usepackage{subfigure}
\begin{document}
\title{Avalanche frontiers in dissipative abelian sandpile model as off-critical SLE(2)}
\author {M. N. Najafi, Saman Moghimi-Araghi, S. Rouhani}
\author{Saman Moghimi-Araghi}
\author{M. N. Najafi, Saman Moghimi-Araghi, S. Rouhani\\
\footnotesize{Physics department, Sharif University of Technology, P.O. Box 
11155-9161, Tehran, Iran}}
\date{}
\maketitle
\begin{abstract}
        Avalanche frontiers in Abelian Sandpile Model (ASM) are random simple curves whose continuum limit is known to be a Schramm-Loewner Evolution (SLE) with diffusivity parameter $\kappa = 2$. In this paper we consider the dissipative ASM and study the statistics of the avalanche and wave frontiers for various rates of dissipation. We examine the scaling behavior of a number of functions such as the correlation length, the exponent of distribution function of loop lengths and gyration radius defined for waves and avalanches. We find that they do scale with the rate of dissipation. Two significant length scales are observed. For length scales much smaller than the correlation length, these curves show properties close to the critical curves and the corresponding diffusivity parameter is nearly the same as the critical limit. We interpret this as the ultra violet (UV) limit where $\kappa = 2$ corresponding to $c=-2$. For length scales much larger than the correlation length we find that the avalanche frontiers tend to Self-Avoiding Walk, the corresponding driving function is proportional to the Brownian motion with the diffusion parameter $\kappa =8/3$ corresponding to a field theory  with $c = 0$. This is the infra red (IR) limit. Correspondingly the central charge decreases from the IR to the UV point.
\end{abstract}
\maketitle
\section{Introduction}

     Sandpile models were introduced by Bak, Tang and Wiesenfeld\cite{BTW} as a prototypical example for a class of models that show Self-Organized Criticality. These models show critical behavior without fine tuning of any external parameter. The abelian structure of the sandpile model was first discovered by D. Dhar, and there after the (generalized) model was named Abelian Sandpile Model (ASM)\cite{Dhar}. Despite its simplicity, ASM has various interesting features and numerous work, analytical and computational, has been done on this model. Among them one can mention different height and cluster probabilities\cite{MajDhar1}, the connection of the model to spanning trees\cite{MajDhar2}, ghost models\cite{MahRuel}, q-state Potts model\cite{SalDup} avalanche distribution\cite{Avalanche}. For a good review see \cite{Dhar2}. Moreover, some of these results are analyzed in the light of a conformal field theory description with central charge $c = -2$\cite{MahRuel,JPR,MRR}. 

Knowing that the model has a conformal field theory description, one may be encouraged to look for geometrical objects that show conformal symmetry. The model has been shown to be related to loop erased random walk (LERW)\cite{MajumLERW}. On the other hand it was realized that LERW belongs to a family of conformally invariant curves called Schramm-Loewner Evolution, SLE$_{\kappa=2}$\cite{Schm}. Schramm-Loewner evolution is a mathematical description of continuum limit of interfaces in 2-dimensional critical statistical mechanical systems. The curves described by the SLE growth process have two essential properties that make them suitable to describe the curves in critical statistical models: conformal invariance and the domain Markov property\cite{Cardy}. It is also shown that SLE$_\kappa$ is related to conformal field theory with central charge $c=(3\kappa-8)(6-\kappa)/2\kappa$\cite{BauBer1}. This suggests that ASM, which has correspondence with a $c=-2$ conformal field theory, should have geometrical objects which are described by SLE$_\kappa$ with $\kappa=2$ or 8. It was numerically shown that the avalanche frontiers in ASM obey the statistics of SLE$_2$\cite{SMDR}.  

Many aspects of the ASM have been studied at criticality, but little work is done on the model out of criticality. Away from the critical point, the conformal invariance of the system breaks and therefore the drift in SLE is expected not to be a simple Wiener process. It is very interesting to see how the drift term properties separates from a Wiener process as we slightly take the system away from the critical point. As the conformal invariance is broken, a length scale, which is a correlation length, should appear in the system. 

 In scales much smaller than the correlation length, but larger than the lattice constant, one expects that the statistical features of the curves fit to the critical case. Since the correlation length could be assumed to diverge at these scales, the conformal symmetry is restored and the evolution of SLE uniformizing map is the same as in the critical case. Therefore we expect to see a Wiener process at small scales with diffusivity parameter $\kappa=2$. On the other hand at scales much larger than the correlation length, the renormalization group (RG) flow may take the system to a new fixed point and we might again see a Wiener process at very large scales but with a new diffusivity\cite{BauBer2}. We suggest that the new critical point corresponds to the Self Avoiding Walk (SAW). 

This paper is organized as follows: In section \ref{ASM} we briefly introduce ASM and off-critical ASM, Section \ref{OCSLE} contains some features of (off-critical) SLE and in section \ref{NSLE} we present some statistical analysis's on waves and avalanches of ASM in the off-critical state and bring our numerical result of application of Schramm-Loewner Evolution on these off-critical avalanches.

\section{Introduction to Abelian Sandpile Model}\label{ASM}
Let us Consider ASM on a two dimensional square lattice $L\times{L}$. To each site a height variable $h_{i}$ is assigned which takes its value from the set $\lbrace{1, 2, 3, 4}\rbrace$. This height variable shows the number of sand grains in the underling site. The dynamics of this model is as follows: in each step, one grain of sand is added to a randomly chosen site $i$, i.e. $h_{i}\rightarrow{h_{i}+1}$.  If the resulting height becomes more than 4, the site topples and looses 4 grains of sand, each of which is transferred to one of four neighbors of the toppled site. As a result, the neighboring sites may become unstable and topple, and in this way a chain of topplings may happen in the system and in this way an avalanche forms. If a boundary site topples, one or two grains of sand (for the sites in the corners of the lattice two grains and for the other boundary sites one grain) will leave the system. The process of toppling continues until the system reaches a stable configuration in which no site has a height greater than 4. The process is then repeated. There are two kind of configurations in ASM: those configurations that may happen once and shall not happen again, which are called transient states, and those which are called recurrent states. In the steady state, no transient state occurs and the recurrent states all occur with the same probability.  It has been shown that the total number of recurrent states is det$\Delta$ where $\Delta$  the discrete Laplacian matrix. For more details see the reference\cite{Dhar2}. This model can be defined on other lattices. For a lattice in which the sites have $z$ neighbors, a site topples if its height exceeds  $z$, the site looses $z$ grains and the height of each of its neighbors will be increased by amount 1. For technical purposes we have done our simulations on triangular lattice in which  $z=6$.

Much work has been done on distribution functions of size, area, gyration radius and etc.\cite{Manna}. It has been shown that the avalanches are not single fractal objects, while waves are\cite{KitLuGrPri}. The waves are constructed in the following way: If, as a result of the addition of a grain to a site $i$, the site becomes unstable, it topples, as do the sites which become unstable as a consequence of the toppling at site $i$. The first wave is the collection of all sites which have toppled given that the initial site is not allowed to topple more than once. It is easy to see that in the wave, the set of toppled sites forms a connected cluster with no voids (untoppled sites fully surrounded by toppled sites), and no site topples more than once. After the first wave is formed completely, if the site $i$ is still unstable, it topples once more to construct the second wave. The process continues until after a number of waves, the site $i$ becomes stable.\\

\textbf{Dissipative ASM}

\vspace*{3mm}
 In the off-critical set up the system has dissipation in the following sense: when the height of a site becomes greater than $x$($>z$), then $z$ grains are transfered to the neighboring sites and $x-z$ grains leave the system. This means that during a toppling in the bulk, the number of sand grains is not conserved anymore, so the system is dissipative. The subject of non-conserved sandpile models has been discussed in \cite{Ghaffari,VespZap,MannaKiss,Tsuchia} and it has been shown that making the model dissipative takes the system to off-criticality. Also, Mahieu and Ruelle showed that this system could be described by massive symplectic fermions\cite{MahRuel}:
\begin{equation}\label{Action}
S=\int{d^{2}z}(\partial\theta\bar{\partial}\bar{\theta}+\frac{m^{2}}{4}\theta\bar{\theta})
\end{equation}
where $m^{2}=x-z$ and $\theta$ and $\bar{\theta}$ are complex Grassmann variables. To study the model at small masses, we should be able to make the mass continuous. It is straightforward to generalize the theory to continuous mass\cite{AziLotMogh}. For simplicity in simulations, we take the mass to be a rational number. The dissipative model is constructed in the following way: Let the threshold beyond which the column of grains becomes unstable be $h_{max}=zn+1$ where $n$ is a arbitrary positive integer number and define $x\equiv\frac{h_{max}}{n}=z+\frac{1}{n}$. Now, during a toppling ($h>h_{max}$), $n$ grains of sand are transferred to each neighboring site and 1 grain leaves the system. With this construction, the effective dissipation will be $m^{2}=x-z=\frac{1}{n}$. The correlation length of such a model is defined as a characteristic length above which the correlation functions of the model become negligible (with respect to the their amount at small distances). At criticality, the correlation length is of order of the size of the system and if the system is taken to be the whole two dimensional plane, it becomes infinite. In the off-critical state however, the correlation length is finite and is a function of the off-critical parameter. This length scale could be observed in many different properties of the model. For example, the Green function of the model will have an exponential decay, rather than a power law. This exponential decay will show up in the correlation function of two site probabilities\cite{MahRuel}. Also there will be a maximum size of an avalanche, say a typical radius, which is of the order of correlation length. For the scales much smaller than this length scale, again a critical behavior is observed, through power law behavior.

\section{Schramm-Loewner Evolution and Off-Criticality}\label{OCSLE}
 Critical statistical systems have special geometrical features. Since in such systems there is no preferred length scale, the system is scale invariant and you can find fractal properties in different objects of the system. In two dimensions, the algebraic aspect of this scale (and conformal) invariance has been well studied in terms of conformal field theories. However the geometric aspects of conformal symmetry is not clear in this approach. Schramm-Loewnwer Evolution is a new approach focused on the geometrical properties of a single curve existing in the model, conditioned to start at the boundary, in the background of all the others. In fact, in this approach, instead of studying the local observables, we focus on the interfaces within two dimensional models. These domain walls are some non-intersecting curves which reflect the essential properties of the system in question. They are supposed to have two properties: conformal invariance and the domain Markov property\cite{Cardy}. Schramm-Loewner Evolution is the candidate to analyze these random curves by classifying them to the one-parameter classes (SLE$_{\kappa}$).

SLE$_{\kappa}$ is a growth process defined via conformal maps, $g_{t}(z)$, which are solutions of Loewner's equation:
\begin{equation}
\partial_{t}g_{t}(z)=\frac{2}{g_{t}(z)-\xi_{t}}
\end{equation}
where the initial condition is $g_{t}(z)=z$  and $\xi_{t}$ is a continuous real valued function and is shown to be proportional to the Brownian motion ($\xi_t=\sqrt{\kappa}B_t$ where $B_{t}$ is the one dimensional Brownian motion and $\kappa$ is the diffusivity constant) if the curves have the two above properties. For fixed $z$, $g_{t}(z)$ is well-defined up to time $\tau_{z}$ for which $g_{\tau_z}(z)=\xi_{t}$. Let us denote the upper half plane by  $H$ and  $\gamma_{t}$ as the SLE trace i.e.  $\gamma_{t}=\lbrace z\in H:\tau_{z}\leq t \rbrace$ and the hull $K_{t}=\overline{\lbrace z\in H:\tau_{z}\leq t \rbrace}$. The complement $H_{t}:=H\backslash{K_{t}}$ is simply connected and $g_{t}(z)$ is the  conformal mapping $H_{t}\rightarrow{H}$ with $g_{t}(z)=z+\frac{2t}{z}+O(\frac{1}{z^{2}})$ as $z\rightarrow{\infty}$ that is known as hydrodynamical normalization. One can retrieve the SLE trace by $\gamma_{t}=\lbrace\lim_{\epsilon\downarrow{0}}g_{s}^{-1}(\xi_{s}+i\epsilon),s\leq t\rbrace$. There are phases for these curves, $2\leq\kappa\leq{4}$ the trace is non-self-intersecting and it does not hit the real axis; $k_{t}=\gamma_{t}$, but for $4<\kappa\leq{8}$, the trace touches it self and the real axis so that a typical point is surely swallowed as $t\rightarrow\infty$ and $K_{t}\neq\gamma_{t}$. It has also been shown that SLE$_\kappa$ has correspondence with CFT with central charge $c_{\kappa}=\frac{(6-\kappa)(3\kappa-8)}{2\kappa}$ and at the point where the SLE curve starts a boundary changing operator with weight $h_{1;2}=\frac{6-\kappa}{2\kappa}$ is sitting\cite{BauBer1}.
\newline 
\newline \textbf{SLE Out of criticality}; Tuning some external parameters, it is possible to take a system off its criticality. In this case the conformal invariance of the system is broken and the system's correlation length $\zeta$\footnote{We have called the correlation length $\zeta$ so that it will not be confused with the drive function $\xi_t$} is finite and will play a crucial role in statistical properties of the system. The question will then be that if we break the conformal symmetry, what would be the effect on Schramm-Loewner Evolution. In other words, when the curves do not have conformal symmetry, then there is no guarantee that the drive function $\xi_t$ be proportional to a Brownian motion, and its statistics have to change as we go away from criticality. However it is expected that at small scales (comparing with correlation length)  i.e. in the ultraviolet regime (UV),  the deviations from criticality are small. This means that the interface should look locally like the critical interface and over short time periods, the off-critical driving function $\xi_{t}^{\zeta}$ should not be much different from the critical state. On the other hand, at large scales (comparing with $\zeta$), i.e. in the infrared regime, it is possible that the RG flow take the system to a new fixed point with conformal symmetry, and the interfaces look like another SLE with a new $\kappa=\kappa_{IR}$. If this is the case, the overall behavior of the curve could be explained as follows: the growing curve forms some bulbs at small scales, with properties close to the original critical system (with diffusivity $\kappa_{UV}$).  When the linear size of the bulb becomes comparable with the correlation length, the curve travels to a new region to form a new bulb. When we integrate out the small scale objects to reach the large scale properties, the bulbs which is formed by SLE trace, may be seen as points that the SLE trace crosses with the new diffusivity ($\kappa_{IR}$). One example is the Ising model. At criticality we have $\kappa=3$, but if the temperature is raised above the critical point, renormalization group arguments indicate that at large scale the interfaces of the model look like the interfaces at infinite temperature i.e. percolation with $\kappa_{IR}=6$\cite{BauBer2}. It is interesting to see if this happen in some other models, such as ASM i.e. $c=-2$ theory. The question could be traced in the context of perturbed CFT's\cite{Zamol}. The special case we are interested in, is the perturbed Logarithmic CFT, as it is known the $c=-2$ is actually a logarithmic CFT. In \cite{RajabRouh} various deformations to a LCFT is considered and it is discussed that the IR fixed point of the massive perturbation of $c=-2$ theory is $c=0$ conformal field theory. They proposed that this theory could be critical percolation in which the domain-walls separating black and white sites are SLE$_6$ (whose dual SLE is SLE$_{8/3}$ (SAW)).

\section{Off-critical ASM and  Schramm-Loewner Evolution, Numerical Data and Results}\label{NSLE}
In this section we present some numerical results on off-critical ASM and its relation with SLE. In the following subsection we focus on how off-criticality affect basic properties of ASM and then we apply SLE to the model to see what happens in presence of dissipation.

\subsection{Off-critical ASM}
To have a feeling about the off-critical ASM, let us study the above mentioned properties of the off-critical model: Green function, two-point functions and the statistics of waves and avalanches of ASM and its dependence on the dissipation. 

\textbf{Green Function.}  To begin, it would be proper to see how the Green function changes as dissipation is introduced to the system. The Green function of ASM is defined as follows\cite{Dhar2}: $G(\vert i-j\vert)$ is the number of topplings occurring in the site $j$ (up to a normalization factor) if you add a grain of sand to the site $i$. It is easy to show that the Green function is just the inverse of the matrix $\Delta$, therefore in two dimensions, asymptotically it behaves as a logarithmic function. The same matrix $\Delta$ appears in the action of the field theory associated with the model whose continuous limit is the free massless Grassmann action if there is no dissipation. Now if we add dissipation, the Grassmann field will become massive (the action in Eq[\ref{Action}]. Therefore in this context, the new Green function is the solution of the equation $(\frac{1}{r}\partial_{r}[r\partial_{r}]-m^{2})G(r,\acute{r})=\delta(r-\acute{r})$ which is the modified Bessel function of 2nd kind. The asymptotic behavior of this function is $\exp(-m r) /\sqrt{r}$ which means that a correlation length $\zeta\sim 1/m$ is introduced to the system. At scales much less than the correlation length, the Green function revive its critical logarithmic behavior.  

\begin{figure}
\begin{picture}(110,130)(0,0)
\includegraphics{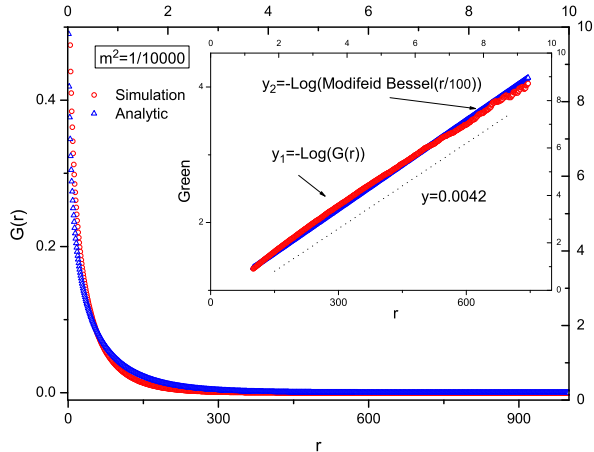}
\end{picture}
\caption{The Green function obtained from simulation and the analytic solution. Inner graph shows the asymptotic behavior of the simulated and analytic Green function $G(r)$ vs large distances for $m^{2}=1/10000$.}
\label{Green}
\end{figure}

On the other hand one can find the Green function numerically from its definition stated above. In FIG[\ref{Green}] the modified Bessel function and the simulated result are sketched in a single plot. In the limit where the distance is much larger than the lattice spacing, the two curves coincide. The inset graph shows this property. For small values of $r$, yet much larger than the lattice spacing, the simulated Green function shows a logarithmic behavior as expected. 

\textbf{Correlation Function} One may ask what is the probability that two distinct sites, distance $r$ apart have heights equal to one, $P(h_0=1,h_r=1)$. The quantity $P(h_0=1,h_r=1)-[P(1)]^2$ then expresses the correlation function of height-one sites ($P(1)$ is the probability of finding a site with height 1). There is a straightforward way to calculate this quantity based on a test called the burning test\cite{MajDhar1}. It turns out that in the expression of the correlation function, the derivatives of Green function appear. This may also be seen in the field theoretical approach, where it has been shown that the field associated with height 1 is proportional to $\bar{\partial}\theta\partial\bar{\theta}+ \partial\theta\bar{\partial}\bar{\theta}$\cite{MahRuel,MoghRajRouh}. In the presence of dissipation, the Green function has an exponential decay, which will be reflected in the behavior of the correlation function. The off-critical two-, three- and four-point functions have been studied in \cite{MahRuel}.

\textbf{Statistics of Waves and Avalanches}. The waves of ASM, as shown in \cite{KitLuGrPri}, are single-fractal objects, whereas the avalanches are multi-fractal. In simulations, we have considered small dissipation rates i.e. $10^{-2}\leq m^2\leq 10^{-4}$, on the triangular lattice which is more symmetric with respect to the square lattice. Also to identify domain walls, we have to consider the honeycomb lattice (the dual of the main lattice) which is more appropriate. The measurements are averaged over $10^{5}$ independent samples on the lattice of size 2048$\times$2048.

\begin{figure}
\begin{picture}(100,175)(0,0)
\includegraphics{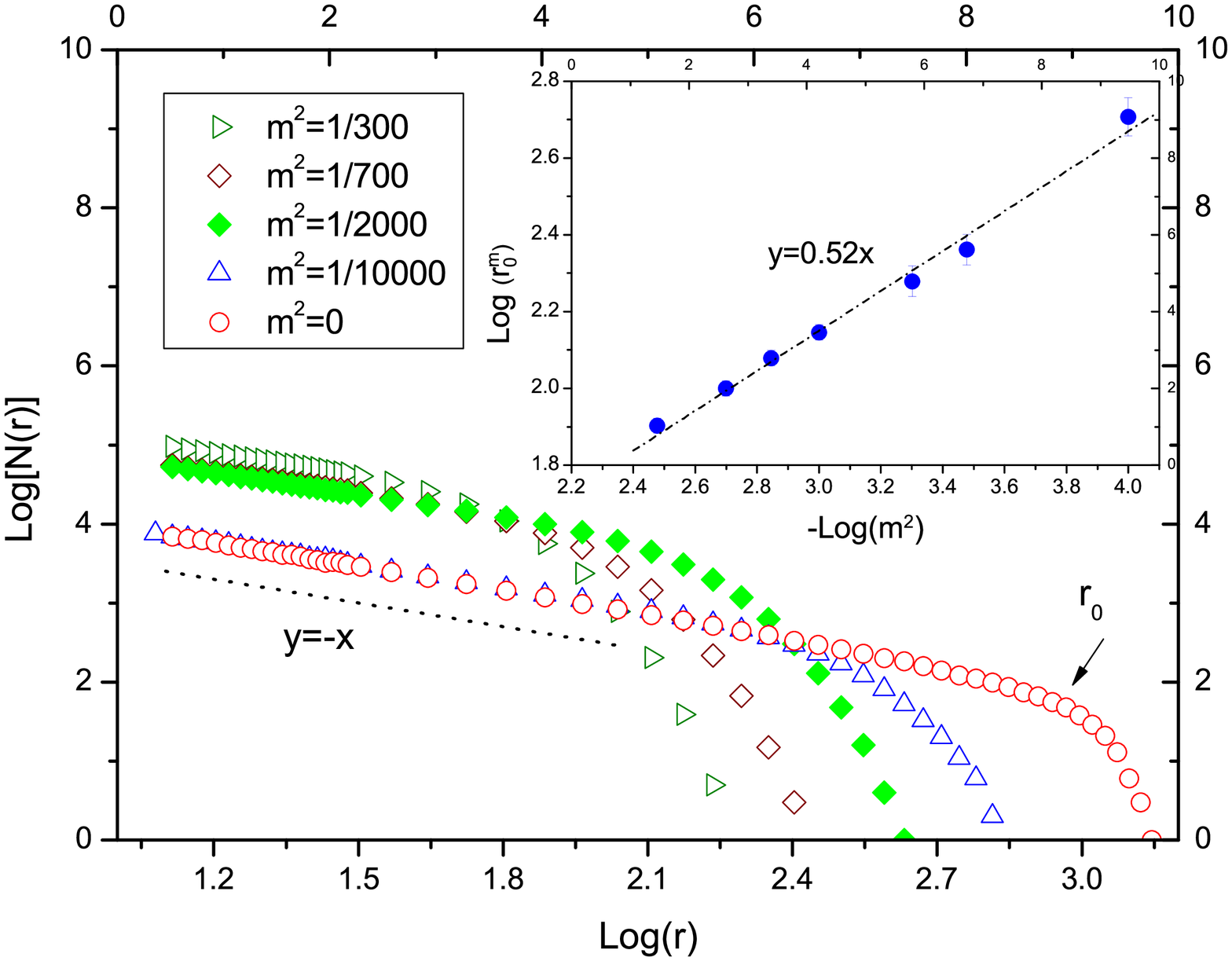}
\includegraphics{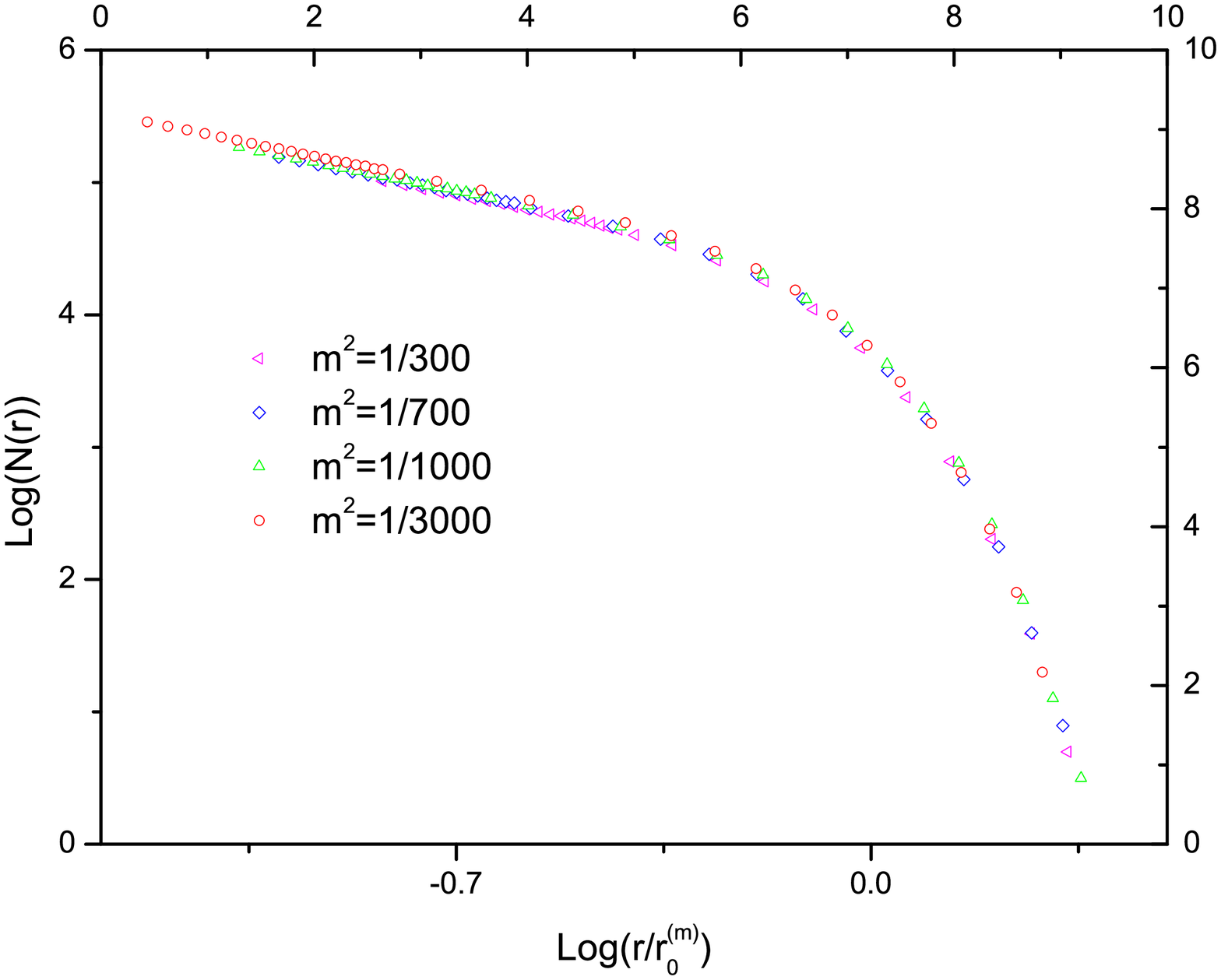}
\put(115,5){(a)}
\put(320,5){(b)}
\end{picture}
\caption{(a) The behavior of $\log(N(r))$ versus gyration radius Log(r), for various dissipation rates for waves. The inner graph shows scaling of their slopes in terms of mass. (b) Log-Log plot of $N(r)$ versus rescaled distance $r/r_{0}^{(m)}$}
\label{gyration}
\end{figure}

\begin{figure}
\begin{picture}(100,170)(0,0)
\includegraphics{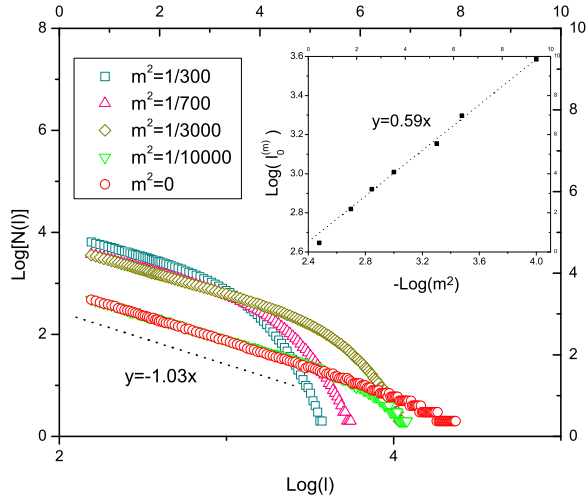}
\end{picture}
\caption{Log-Log plot of loop length distributions $N(l)$ versus $l$ of waves. The inner graph shows scaling of their slope in terms of mass.}
\label{loops}
\end{figure}

FIG. [\ref{gyration}a] shows the behavior of probability distribution of waves, $N(r)$, versus gyration radius ($r$) in a Log-Log plot. The simulation is done with various dissipation rates over a range of about two orders of magnitude. For small radii of gyration, the Log-Log plot is linear, that is $N(r)\sim r^{-\tau_{r}}$ with $\tau_{r}\simeq 1$.  Increasing the dissipation rate causes the linear part of the plot become smaller. In fact for generic $m$ and $r$ we found the data are best fitted with:
\begin{equation}
N(r)=N_{0} r^{-\tau_{r}}\exp[-(r/r_{0}^{(m)})^{a}]
\label{fit}
\end{equation}
in which $N_{0}$ is a normalization factor and $\tau_{r}\simeq 1$ is the same as the critical exponent of $N(r)$ and $r_{0}^{(m)}$ is the point where the linear behavior completely disappear and the plot rapidly falls and $a$ is a parameter to be obtained by fitting data. The best fitted value of $a$ is $2.8\pm 0.1$. $r_{0}^{(m)}$ should be of the order of correlation length. To obtain this quantity from Eq [\ref{fit}], we have sketched in fig[\ref{gyration}b] the Log-Log plot of probability distribution $N(r)$ versus rescaled distance ($r/r_{0}^{(m)}$) in such a way that the curves corresponding to the various masses coincide. The obtained quantities have been plotted in the inner graph of fig[\ref{gyration}a]. It is explicit in this figure that $r_{0}^{(m)}\sim{m^{2\alpha}}$ where $\alpha=0.52\pm{0.04}$; in other words we have $r_{0}^{(m)}\sim{\frac{1}{m}}$ which one expects from the general properties of correlation length. The simulation of avalanches also show the same result which firms this claim. The same behavior is seen for the probability distribution of loop lengths of waves, $N(l)$. In Log-Log plot of $N(l)$ versus $l$ for each dissipation in FIG[\ref{loops}], a characteristic loop length $l_{0}^{(m)}$ emerges where the linear character of the graph is transformed to a rapidly falling behavior. In the inner graph we have shown $l_{0}^{(m)}$ versus $m^{2}$. The scaling relation between $l_{0}^{(m)}$ and $r_{0}^{(m)}$ is $l_{0}^{(m)}\sim{(r_{0}^{(m)})^{D_{f}}}$. If we ignore the change of fractal dimension due to dissipation and take it equal to 1.25, we have $l_{0}^{(m)}\sim{(r_{0}^{(m)})^{D_{f}}}=(m^{-2\alpha})^{D_{f}}=m^{-2\beta}\rightarrow{\beta=\alpha{D_{f}}}\simeq{0.65}$. From the figure we see that $\beta=0.59\pm{0.04}$. The difference is due to the change of fractal dimension for various dissipation rates. In fact we have $\delta \beta=D_{f}\delta \alpha+\alpha\delta D_{f}$, setting $\delta \alpha=0.04$ and $\delta D_{f}=0.05$ (corresponding to its maximum change for various masses) we obtain $\delta \beta\simeq 0.08$ which explains the difference. We would like to emphasize that the finite size scaling effects are not seen as we have considered a relatively large system size ($2048\times 2048$) and the dissipation is not that small ($r_0^{m}\ll 2048$  for all masses). 

Now we focus on a more geometric property, the fractal dimension of the wave frontiers. For the scales much smaller than the correlation length, one expects that the system have a well-defined fractal dimension. We have computed the fractal dimension of waves defined as $l\sim r^{D_{f}}$ and found that for finite masses, there is a slight deviation from the critical fractal dimension \footnote{Note that surely for the lengths comparable with or larger than the correlation length, fractal dimension does not make sense.}. The interesting feature is that this quantity does scale with the dissipation with $D_{f}^{m}-D_{f}^{0}\sim m^{2\gamma_{d}}$ with $\gamma_{d}=0.25\pm 0.02$. Table[1] and Table[2] show the full information of various exponents of the waves and avalanches. Note that the exponents $\tau_{r}=0.95\mp 0.05$ and $\tau_{l}=1.02\mp 0.05$ ($N(l)\sim l^{-\tau_{l}}$) and $\tau_{s}=1.00\mp 0.05$ ($N(s)\sim s^{-\tau_{s}}$) do not change with mass for waves, but their change is relevant for avalanches. More precisely, the probability distribution $N^{AV}(r)$ or the avalanche gyration radius ($r$) shows the same dependence as  Eq [\ref{fit}] but with dissipation dependent exponent $\tau^{Av,m}_{r}$ with the scaling relation $(\tau^{AV,0}_{r}-\tau^{AV,m}_{r})\sim (m^{2})^{\gamma_{r}}$ with $\gamma_{r}=0.50\pm 0.05$. The parameter $a$ is the same as the value for waves i.e. $a=2.8\pm 0.1$.



\begin{table}[h]
\begin{center}
\begin{tabular}{|c|c|c|c|c|c|c|}
\hline $m^{2}$ & $r_{cut}$ & $l_{cut}$ & $\tau_{r}(\pm{0.05})$ & $\tau_{l}(\pm{0.05})$ & $\tau_{s}(\pm{0.05})$ & $D_{f}^{m}(\pm{0.03})$ \\ 
\hline $\frac{1}{300}$ & $80\pm{2}$ & $443\pm{25}$ & $0.90$ & $1.02$ & $1.00$ & $1.3$ \\ 
\hline $\frac{1}{700}$ & $120\pm{4}$ & $835\pm{20}$ & $0.93$ &  $1.03$ & $1.02$ &$1.29$ \\ 
\hline $\frac{1}{3000}$ & $230\pm{5}$ & $1930\pm{106}$ & $0.93$ & $1.01$ & $1.01$ &$1.27$ \\ 
\hline $\frac{1}{10000}$ & $510\pm{12}$ & $3850\pm{120}$ & $0.95$ & $1.01$ & $1.03$ &$1.24$ \\ 
\hline
\end{tabular}
\caption{The $r_{cut}$ and $l_{cut}$ and the exponents of gyration radius $N(r)\sim{r^{\tau_{r}}}$, the fractal dimension $\langle Log(l)\rangle = D_{f}^{m}\langle Log(r)\rangle$ in terms of dissipation strength for waves ($L=2048$).}
\end{center}
\label{table1}
\end{table}

\begin{table}[h]
\begin{center}
\begin{tabular}{|c|c|c|c|c|c|c|}
\hline $m^{2}$ & $r_{cut}$ & $l_{cut}$ & $\tau_{r}(\pm{0.02})$ & $\tau_{l}(\pm{0.04})$ & $\tau_{s}(\pm{0.02})$ & $D_{f}^{m}$ \\ 
\hline $\frac{1}{300}$ & $90\pm{2}$ & $970\pm{24}$ & $1.002$ & $1.035$ & $1.01$ & $1.28\pm{0.01}$\\ 
\hline $\frac{1}{700}$ & $140\pm{6}$ & $1492\pm{32}$ & $1.095$ & $1.084$ & $1.07$ &$1.27\pm{0.01}$\\ 
\hline $\frac{1}{3000}$ & $300\pm{12}$ & $3971\pm{150}$ & $1.225$ & $1.162$ & $1.11$ & $1.26\pm{0.01}$\\ 
\hline $\frac{1}{10000}$ & $540\pm{15}$ & $--$ & $1.29$ & $1.25$ & $1.14$ & $1.25\pm{0.01}$\\ 
\hline
\end{tabular}
\caption{The $r_{cut}$ and $l_{cut}$ and the exponents of the distribution of gyration radius $N(r)\sim{r^{\tau_{r}}}$, the distribution of loop lengths $N(l)\sim{l^{\tau_{l}}}$, the distribution of loop areas $N(s)\sim{s^{\tau_{s}}}$, in terms of dissipation strength for avalanches ($L=2048$).}
\end{center}
\label{table2}
\end{table}

\subsection{Off-critical ASM and relation with SLE}

\begin{figure}
\begin{picture}(100,140)(0,0)
\includegraphics{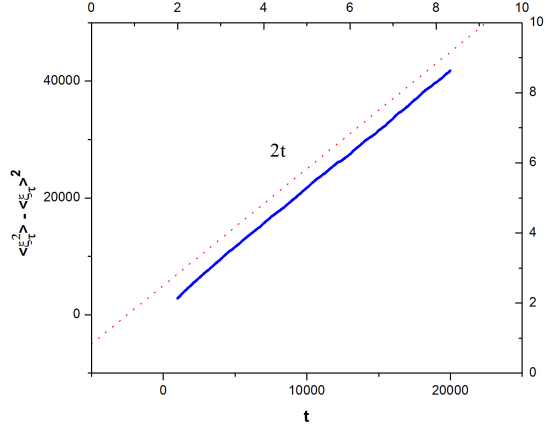}
\end{picture}
\caption{The averaged squared $\xi_{t}$ versus $t$ shows the diffusivity $\kappa=2.0\pm{0.1}$.}
\label{kappa2}
\end{figure}

In this subsection we present some numerical results obtained by applying SLE to the critical and off-critical Abelian sandpile model. The frontiers of waves form the set of loops with discrete points. We have used the algorithm introduced in \cite{BBCF} to obtain the statistics of driving function $\xi_t$. FIG. [\ref{kappa2}] contains the graph $\langle\xi_{t}^{2}\rangle-\langle\xi_{t}\rangle^{2}$ versus $t$ for the critical case. It is observed that $\langle\xi_{t}\rangle\simeq{0}$ and as it is clear in the graph $\xi_{t}$ has the expected variance $(\langle\xi_{t}^{2}\rangle-\langle\xi_{t}\rangle^{2})=\kappa{t}$ with $\kappa=2.0\pm{0.1}$. In the interval $0<t<1000$ the behavior of the graph is a little different due to finite size of lattice spacing, therefore this part of the graph has been ignored. The same effect is observed in the off-critical case, therefore we have neglected this part in all of the following similar graphs. It is worth adding that we do not see the effect of the system size, as we have not reached the 'time' scales in which the typical size of the SLE curves becomes comparable with the system size.
\begin{figure}
\begin{picture}(100,175)(0,0)
\includegraphics{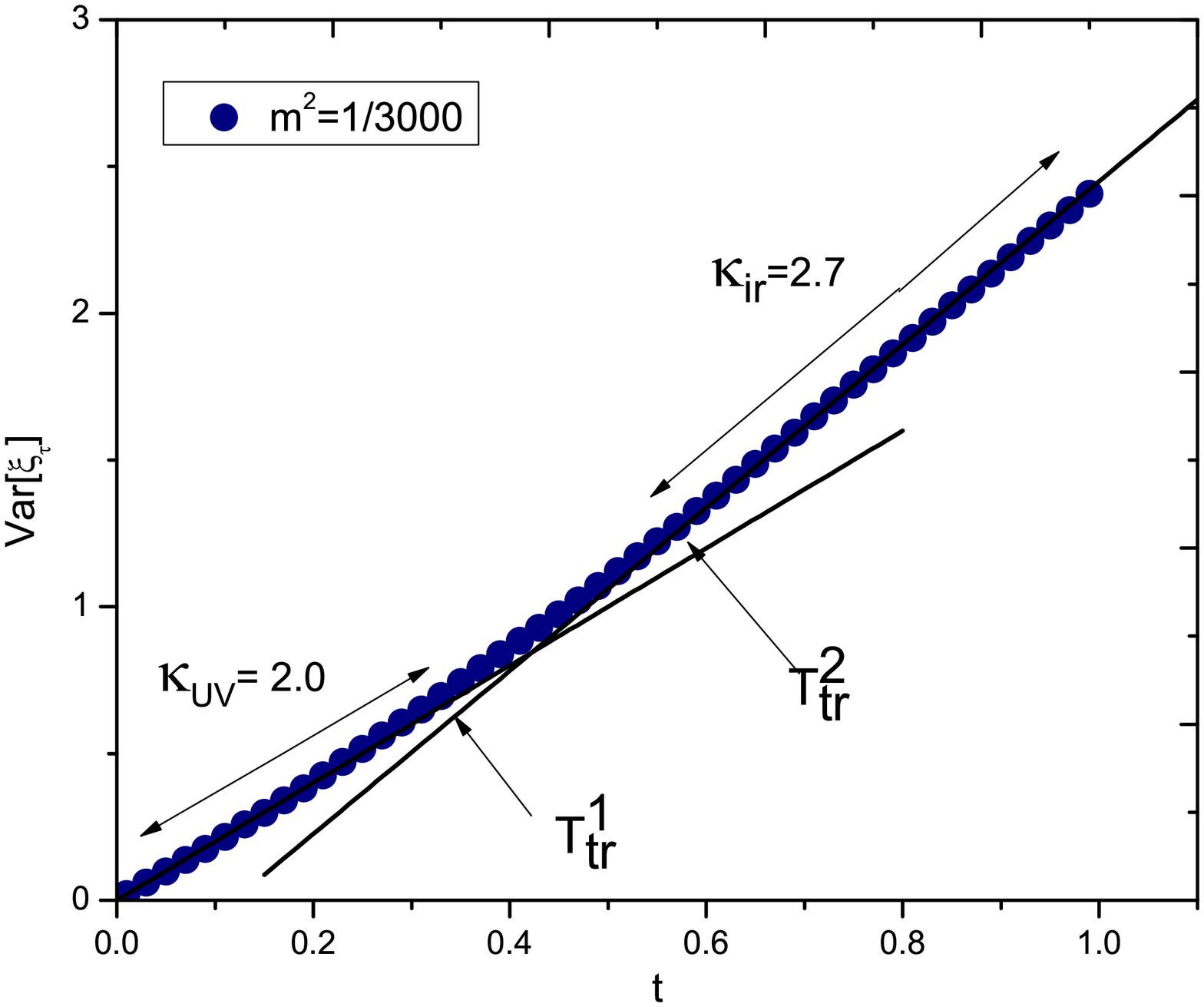}
\includegraphics{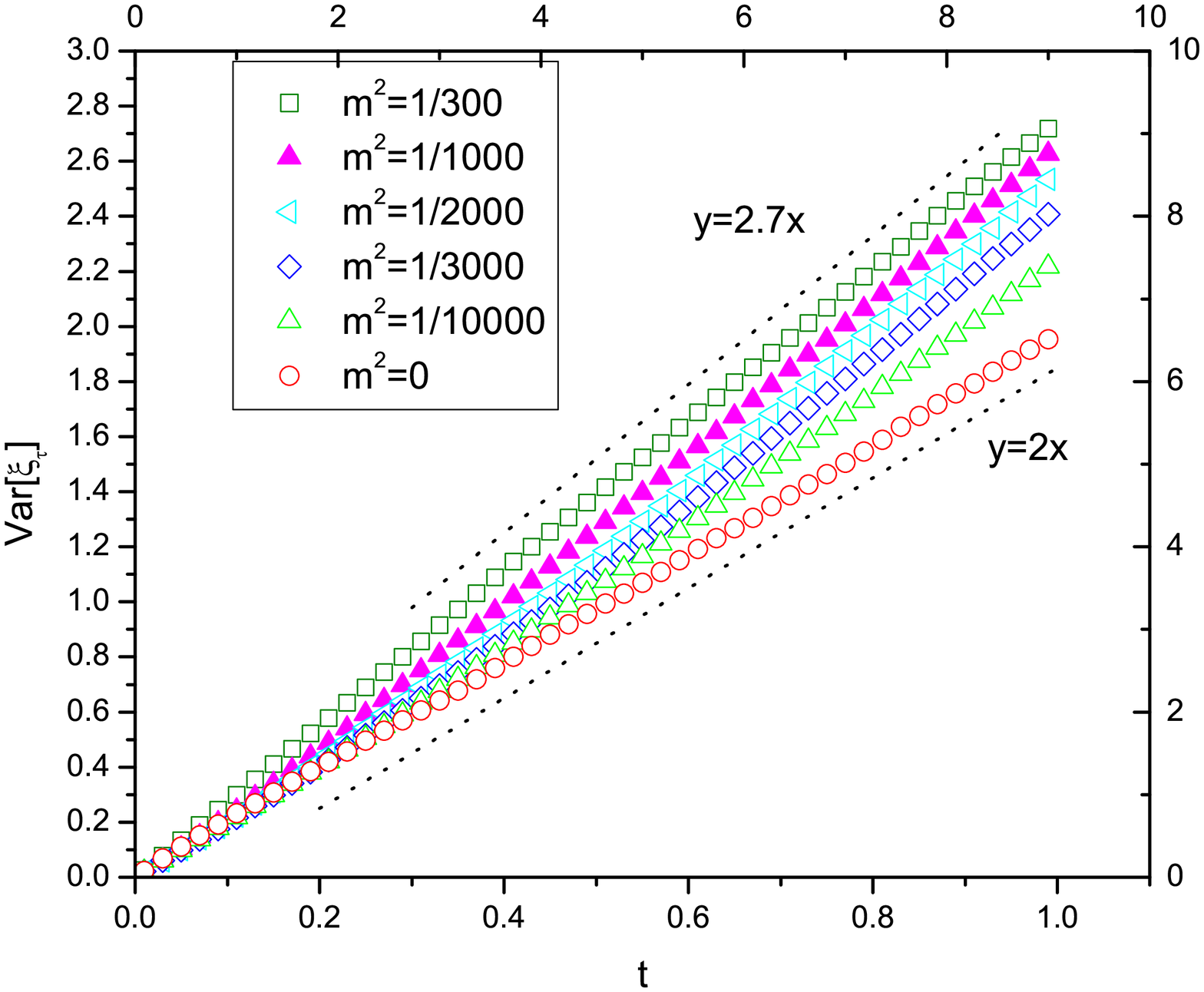}
\put(89,00){(a)}
\put(340,00){(b)}
\end{picture}
\caption{(a) The typical behavior of var[$\xi_{t}$] vs $t$. (b)  var[$\xi_{t}$] versus $t$ for various rates of dissipation}
\label{sample2}
\end{figure}

Now we add dissipation. As stated in the introduction, at scales much smaller than the correlation length, the deviation from criticality should be small, i.e. the interface should look locally like the critical interface. At large scales, i.e. much larger than the correlation length, the interface may look like another SLE with the different $\kappa$. The typical form of the driving function for massive SLE is shown in FIG. [\ref{sample2}(a)]. The graph shows $\langle\xi_{t}^{2}\rangle-\langle\xi_{t}\rangle^{2}$ versus $t$ for $m^{2}=\frac{1}{3000}$. It has two linear regimes and there is a crossover region in between. In the first linear regime  (called the UV regime), the length scale of the growing interface is much smaller than the correlation length of the model. The obtained driving function in this region is not much different from the critical one, as expected (in this graph, $\kappa_{UV}=2.0\pm{0.07}$). In the second linear regime (called the IR regime) the graph becomes linear once again with a different slope $\kappa_{IR}$ (here $\kappa_{IR}=2.7\pm{0.03}$). In FIG[\ref{sample2}(b)] we have sketched  $\langle\xi_{t}^{2}\rangle-\langle\xi_{t}\rangle^{2}$ versus $t$ for various dissipation rates. There are two transition points:  1)  The transition from UV regime to the "cross over region" taking place at $t_{tr}^{(1)}$ and 2) The transition from the "cross over region" to the IR regime taking place at $t_{tr}^{(2)}$. If the above explanation is true, one expects that decreasing dissipation rate, $t_{tr}^{(1)}$ and $t_{tr}^{(2)}$ should  increase. Table [3] shows the value of these transition time for a few different dissipation rates in which the effect stated above is clearly seen.  

\begin{figure}
\begin{picture}(100,170)(0,0)
\includegraphics{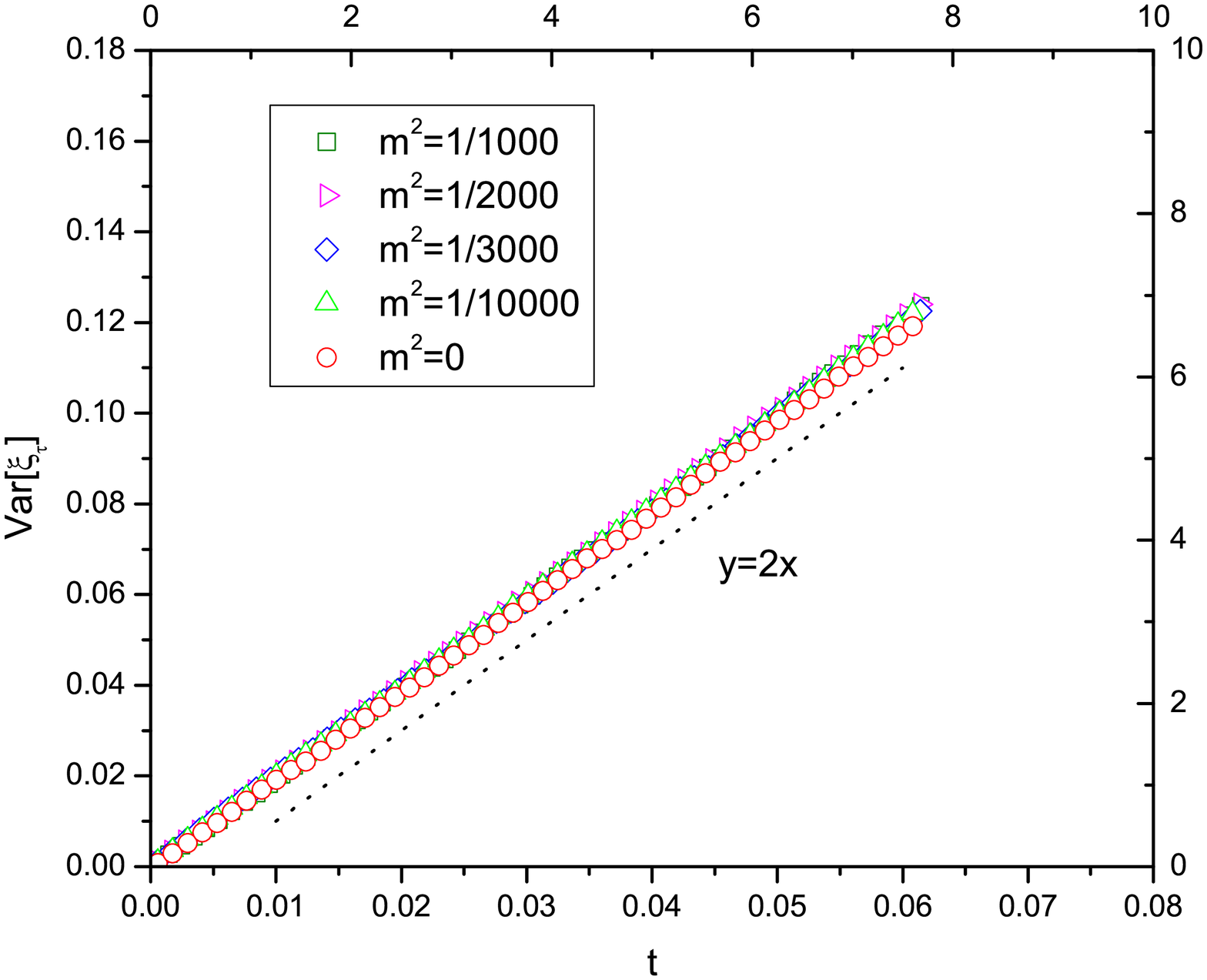}
\includegraphics{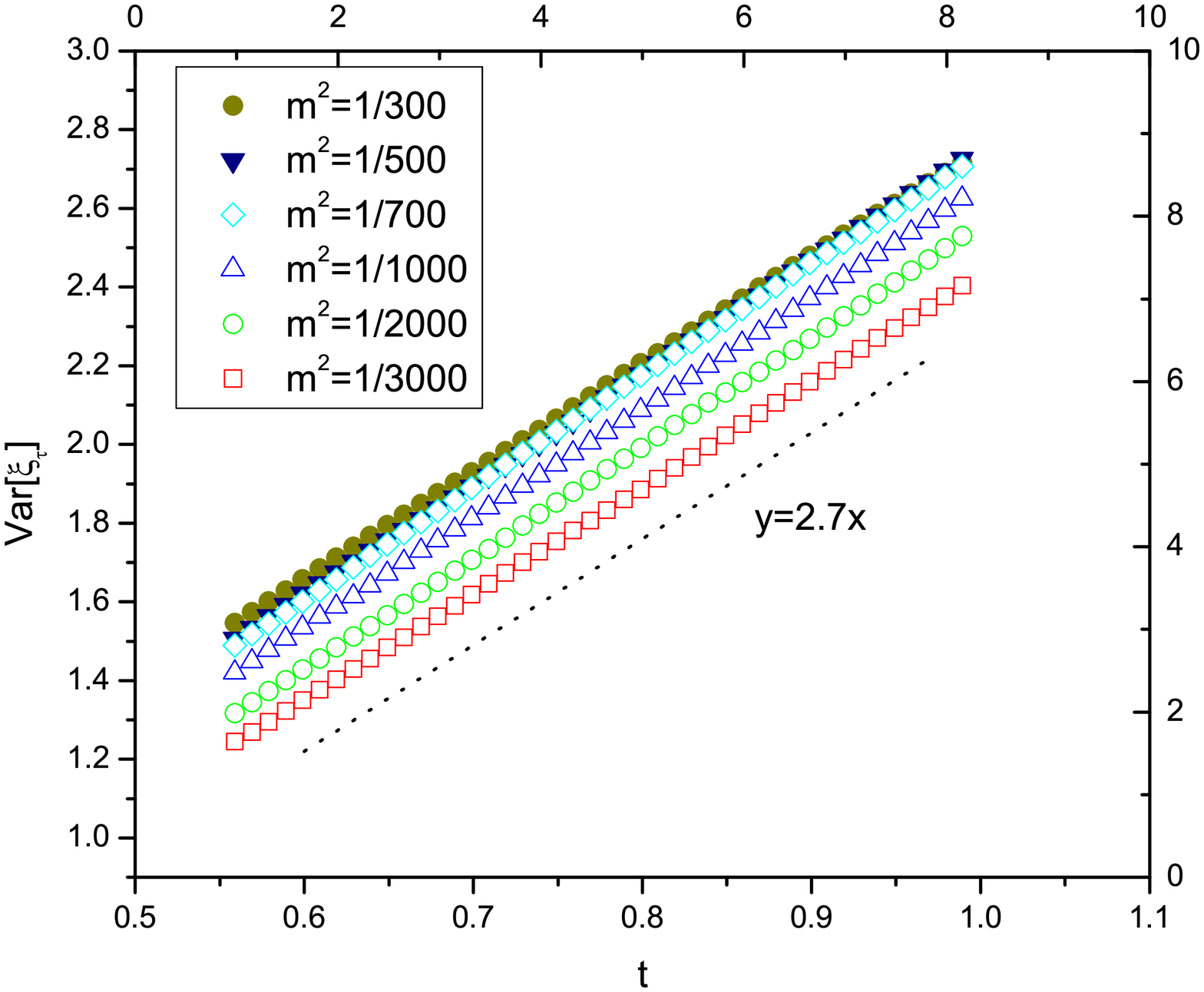}
\put(85,-0){(a)}
\put(340,-0){(b)}
\end{picture}
\caption{(a) $\kappa_{UV}$ for various dissipation rates in UV regime. (b) $\kappa_{IR}$ for various dissipation rates in IR regime.}
\label{kappaa}
\end{figure}

We investigated the behavior of the UV and IR parts of the graphs away from the crossover region more carefully. The curves have a linear behavior with good accuracy in each region and the slopes in IR and UV parts of the graphs has little dependence on the mass parameter. In FIG [\ref{kappaa}(a),\ref{kappaa}(b)], UV and IR regions are sketched separately for various dissipation parameters. In the UV linear regime, all the curves coincide and have the same slope $\kappa_{UV}\simeq 2.0$ within the error bars we have. In the IR part however, the curves are separated (which is due to the fact that the crossover-region time interval is different for different dissipation rates) but have the same slope. Within the error bars we have, all the curves could be fitted to a slope equal to $\kappa_{IR}\simeq 2.7$. This diffusivity does actually coincide with the diffusivity of the SAW which is $\kappa=8/3\simeq 2.67$. So the interesting consequence of this argument is that the IR limit of massive ASM seems to be in the universality class of self-avoiding walk whose continuum limit corresponds to the $c=0$ conformal field theory. Observing just a slope is not enough for such a claim. However, we have some other supports. Rajabpur and Rouhani have studied off-critical LCFT's in \cite{RajabRouh} and have considered the effect of adding relevant operators to an LCFT. Using Zamolodchikov's $c$-theorem they have derived the central charge of the IR limit of such theories. They have considered the special case of $c=-2$ LCFT, the most studied LCFT, and have added the mass term as a relevant operator to the action. They have derived that the IR limit of such thaery has the central charge $c=0$.

The other point to be clarified is that one may argue that there exists a length scale in the model, as stated in the previous subsection. Then how it is possible to have a critical behavior in such a system. We can say that the length scales corrsponding to the second region is much larger that the length scale of the theory. As an example in the case of $m^2=1/1000$, the correlation length is about 100, while the length scale corresponding to $t>t_{tr}^{(2)}$ is about an order of magnitude larger than that. Therefore the correlation length is possibly now the smallest scale of the IR theory.

 \begin{table}
\begin{center}
\begin{tabular}{|c|c|c|}
\hline $m^{2}$ & $t_{tr}^{(1)}$ & $t_{tr}^{(2)}$\\ 
\hline $1/1000$ & $0.062\pm 0.006$ & $0.415\pm{0.035}$\\ 
\hline $1/2000$ & $0.112\pm 0.004$ & $0.502\pm{0.02}$\\
\hline $1/3000$ & $0.165\pm 0.015$ & $0.55\pm{0.02}$\\ 
\hline $1/10000$ & $0.21\pm 0.01$ & $---$\\  
\hline
\end{tabular}
\caption{The explicit values of $t_{tr}^{(1)}$ and $t_{tr}^{(2)}$ for various rates of dissipation.} 
\end{center}
\end{table}

\section{Conclusion}
In this paper, we analyzed the statistics of wave and avalanche frontiers of dissipative ASM, arguing that this must correspond to off-crtical SLE. We observed that there is a scale above which the purely power law behavior of the distribution function of gyration radius disappears ( $r_{0}^{(m)}$ ). This quantity has the same scaling behavior versus dissipation as the expected behavior of the correlation length (Note that the correlation length of the dissipative ASM is the inverse of the mass introduced in the massive ghost action).
Using  SLE technique, we found numerically that for the scales much smaller than the correlation length, the curves are conformally invariant with the same properties as the critical ASM with nearly the same diffusivity ($\kappa=2.0 \pm0.1$). For larger scales, the curves acquire the new diffusivity constant $\kappa_{IR}\simeq 8/3$ and thus the massive ASM tends to SAW. The transition point to this infra red region is mass dependent.

\end{document}